\documentclass[12pt]{article}
\usepackage{a4}
\usepackage{epsf}
\begin{document}
\title{\bf Once More on a Colour Ferromagnetic Vacuum State
            at Finite Temperature}

\author{Vladimir Skalozub\thanks{e-mail: Skalozub@ff.dsu.dp.ua}\\
Dniepropetrovsk State University, 320625 Dniepropetrovsk,  Ukraine \\
and \\
Michael Bordag\thanks{Michael.Bordag@itp.uni-leipzig.de}\\
 Institute for Theoretical Physics, University of Leipzig, \\
Augustusplatz 10, 04109 Leipzig,  Germany}
\maketitle

\begin{abstract}

The spontaneous vacuum magnetization at finite temperature is
investigated in $SU(2)$ gluodynamics within a consistent effective
potential approach including the one-loop and the correlation
correction contributions. To evaluate the latter ones the high
temperature limits of the polarization operators of charged and
neutral gluon fields in a covariantly constant magnetic field and at
high temperature are calculated.The radiation mass squared of charged
gluons is found to be positive. It is shown that the ferromagnetic
vacuum state having a field strength of order $(gH)^{1/2} \sim g^{4/3}
T$ is spontaneously generated at high temperature. The vacuum
stability and some applications of the results obtained are discussed.
\end{abstract}

\section{Introduction}
The problem on generation of magnetic fields in nonabelian gauge theories
at finite temperature is of great importance for particle physics and
cosmology. Its positive solution, in particular, will give a theoretical basis
for investigations of the QCD vacuum at high temperature and the primordial
magnetic fields in the early universe \cite{Enol}, \cite{Amol}. In literature
different mechanisms of producing the fields are discussed (see recent papers
\cite{Shap1},\cite{Shap2} and references therein). The aim of the present
paper is to investigate in more detail one of them - the spontaneous
magnetization of the vacuum of non-abelian gauge fields at finite temperature.
This problem was studied recently in Refs.\cite{Enol}, \cite{SVZ}, \cite{Ska1}
where the creation of the vacuum magnetic field has been derived. However, a
trusty conclusion about the possibility of this phenomenon as well as
estimations of the value of the condensed field have not been obtained, yet.
In fact, in papers \cite{Enol}, \cite{Ska1} the effective potential (EP) was
calculated in one-loop order. But, as it is known, at finite temperature, $
T$, the correlation corrections to the EP are of great importance. They may in
an essential way influence the properties of a system. In Ref.\cite{SVZ} a
number of the corrections has been taken into account. These authors have
determined that the possibility of the vacuum magnetization depends on the
sign of the next-to-leading term of the polarization operator (PO) of charged
gluons, which has to be calculated in the field $H$ and at $T \ne 0 $.
However, these calculations had not been done there. Besides, in
Refs.\cite{Enol},\cite{SVZ} the temperature mass squared of the unstable mode
present in the spectrum of charged gluons was erroneously identified with the
Debye one having the order $\sim g^2 T^2$. However, the latter mass is
generated for longitudinal modes of gauge fields. The unstable mode
$\epsilon^2 (n = 0, \sigma = 1) = p^2_3 - gH , $ where $ g$ is the gauge
coupling constant, $ n$ is the Landau level number, $\sigma$ is the spin
projection on the field direction, is the transversal one produced by a spin
interaction with the magnetic field. Therefore, the vacuum stabilization
condition at $T \ne 0$ has not been studied in detail.  Moreover, a part of
diagrams describing correlation corrections was missed there. In papers
\cite{Elp},\cite{Pers} the scale of the magnetic field at $T \ne 0$ was
incorrectly estimated and as a consequence the authors came to negative
conclusion about the vacuum magnetization.

To summarize the present time situation we recall that in one-loop 
approximation the Savvidy level of order $ (gH)^{1/2}_{c} \sim
g^2 T $ is generated. But the way of its stabilization and its
existence with the  correlation corrections been taken into account
remain to be investigated.
 
It will be important for what follows to remind recent results on observation
of the gluon magnetic mass in lattice simulations which was found to be of
order $m_{mag} \sim g^2 T$ (as it has been expected from nonperturbative
calculation in quantum field theory \cite{Kal},\cite{Buch}).  The mass screens
magnetic fields at distances $l > l_m \sim (g^2 T)^{-1}$ but inside the space
region $l < l_m$ they may exist. Since the typical values of particle masses
at high temperature are $M \sim gT$, the magnetic fields of order $(gH)^{1/2}
\sim gT$ or at least $\sim g^2 T$ are to be of interest.  They are able to
affect all the processes at high temperatures.
 
In the present paper we investigate the vacuum magnetization at finite
temperature within $SU(2)$ gluodynamics. Considering the Abelian covariantly
constant chromomagnetic field $H^a = \delta^{a3}H = const$ and finite
temperature as a background we calculate the EP containing the one-loop and
the ring diagram contributions of both neutral and charged gluon fields. To
find the latter ones the high temperature limits of the PO of gluon fields at
the magnetic background are also calculated. It will be shown that in the
adopted approximation the Savvidy level with the field strength $(gH)^{1/2}_c
\sim g^{4/3} T$ is produced.  Thus, we come to the conclusion that the
correlation corrections make the field strength stronger as compare to the
value derived in one-loop approximation \cite{SVZ},\cite{Ska1}. This, in
particular, means that although the field is screened for $l > (g^2 T)^{-1}$,
the spectrum of charged particles, being formed at Larmor's radius $r \sim
(gH)^{-1/2}_c \sim (g^{4/3}T)^{-1}$, is located inside this domain $ r
<<l_{m}$ for small $g$. Moreover, since at high temperatures the particle
masses are of order $ \sim gT$ , one has to conclude that a constant
background magnetic field is a good approximation to investigate processes at
high temperature.

The content is as follows. In Sect.2 the one-loop EP (zero and finite
temperature parts) is presented in the form convenient for numerical
investigations. Here we also present a general expression describing
the contribution of the neutral gluon ring diagrams.  In Sect.3 we
calculate the Debye mass of the neutral gauge field at $H,T \ne 0$. In
Sect.4 the same is done for the charged gluon part. Sect.5 contains
calculation of the two-loop diagram taking into account the
self-interaction of charged gluon in the vacuum.  The investigation of
the vacuum magnetization and its stability as well as the discussion
of the results obtained are presented in Sect. 6. In the  APPENDIX
we adduce the necessary  high temperature asymptotic formulae.

\section{Basic Formulae and  General Considerations}

Let us consider the pure Yang-Mills action 
\begin{equation}\label{act} 
S = -\frac{1}{4} \int d^4 x F^a_{\mu \nu}F_a^{\mu\nu},
\end{equation} where \begin{equation} F^a_{\mu \nu} = \partial_{\mu}
A^a_{\mu} - \partial_{\nu} A^a_{\mu} + g f^{abc} A^b_{\mu} A^c_{\nu}
\end{equation} \nonumber is the field strength. In the $SU(2)$ gauge
theory the Savvidy vacuum state is characterized by the uniform
background classical colour magnetic field \begin{equation} \label{H}
H^a_i = \delta^a_3 \delta_{i3} H, \end{equation} where $H$ is a
constant, $a = 1,2,3 $ is the isotopic index.

To introduce this background field one has to decompose the gauge field
potential as
\begin{equation} \label{A}
A^a_{\mu} = A^{a~ ext}_{\mu} + B^a_{\mu},
\end{equation}
where $A^{a~ ext}_{\mu} = \delta_{\mu 2} \delta^a_3 H x_1$ and $B^a_{\mu}(x)$
is the quantum field. In the EP calculations the background gauge fixing
condition
\begin{equation}
\partial_{\mu} B^a_{\mu} + g f^{abc} A^{b~ext}_{\mu}B^c_{\mu} = 0
\end{equation}
\nonumber 
will be used.
 
The thermodynamic potential of the model is
 \begin{equation} \Omega =
- \frac{1}{\beta} \log  Z , \end{equation} 
\nonumber
and
\begin{equation} \label{TP} Z = Tr \exp (-\beta \cal{H})
 \end{equation}
 is the partition function, $\cal{H}$ is the Hamiltonian of the system, $\beta
 = 1/ T $ is inverse temperature and the trace is calculated over all physical
 states.
 
 For what follows it will be convenient to introduce the ``charged basis'' of
 fields
\begin{equation} \label{bas}
W^{\pm}_{\mu}=\frac{1}{\sqrt{2}}(A^1_{\mu} 
\pm iA^2_{\mu}), \  A_{\mu}= A^3_{\mu}.\end{equation} 
In this basis the problem of the vacuum magnetization is reduced to
calculation and investigation of the vacuum polarization of fields $W^+_{\mu},
W^-_{\mu}$ in the external field $A^{ext}_{\mu}$.

To obtain the EP one has to rewrite eq.(\ref{TP}) as a sum over
quantum states calculated near the nontrivial classical solution
$A^{ext}$.  This standard procedure is described in many papers and
textbooks (see, for example, Refs. \cite{Kap},\cite{Dolj},\cite{Car},
\cite{SVZ}) and the result can be written in the form:
\begin{equation} \label{EPr}
V = - \frac{1}{2} \sum\limits_i Tr log G^{(0)}_i + V^{(2)}(H,T) + \cdots,
\end{equation}
where $i$ marks the type of field $(W^{\pm}, A$ and ghosts),
$G^{(0)}_i$ is the corresponding propagator in the external field
$A^{ext}_{\mu}$. In the present paper to incorporate temperature the
imaginary time formalism will be used. In the specified above
background field, the trace means summation over the discrete
Matsubara frequencies, summation and integration over eigen values of
the quantum fields. The first term in Eq.(\ref{EPr}) corresponds to
the one-loop EP whereas the other ones present the contributions of
two-, three-, etc. loop corrections.

Among these terms there are ones responsible for dominant
contributions of large distances at high temperature - so-called daisy
or ring diagrams (see, for example Ref. \cite{Kap}). This part of the
EP, $V_{ring}(H,T),$ is important in the case when massless states
appear in a system. The ring diagrams of either neutral and charged
gluons have to be calculated when the vacuum magnetization at finite
temperature is investigated. Really, one first must assume that the
field is nonzero, calculate EP $V(H,T)$ and after that check whether
the minimum of it is located at nonzero $H$. On the other hand, if one
investigates problems in the applied external field, the charged
fields become massive with the mass $\sim (gH)^{1/2}$ and have to be
omitted. To find $V_{ring}(H,T)$ the one-loop polarization operators
of charged, $\Pi(H,T),$ and neutral, $\Pi^0(H,T), $ gluons in the
external field and at finite temperature have to be calculated in the
limit of zero momenta. Then $V_{ring}(H,T)$ is given by the series
depicted in figures 1,2. Here, the dashed lines describe the neutral
gluons and the wavy lines represent the charged ones, blobs stand for
the one-loop polarization operators. The diagrams with one blob show the
two-loop terms of the EP, with two blobs - three-loop ones, etc.

\begin{figure}[ht]
\unitlength1cm
\begin{picture}(14,2)
\put(0,-2.5){\epsfxsize=10cm\epsfysize=8cm\epsffile{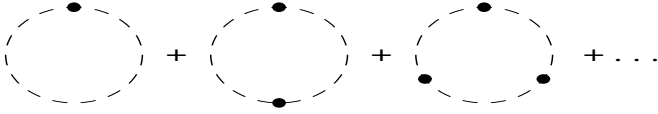}}
\end{picture}
\caption{The neutral gluon ring diagrams giving contribution to the effective
potential.}
\end{figure}
\begin{figure}[ht]
\unitlength1cm
\begin{picture}(14,2)
\put(0,-2.5){\epsfxsize=10cm\epsfysize=8cm\epsffile{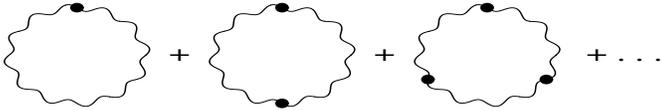}}
\end{picture}
\caption{The charged gluon ring diagrams giving contribution to the
effective potential.}
\end{figure}

The standard procedure to incorporate these contribution is to
substitute $G^{[0]}_i$ in Eq. (\ref{EPr}) by the inverse full
propagator $G^{-1}_i = [G_i^{(0)}]^{-1} + \Pi(H,T) $ \cite{Kap},
\cite{Car}. In the case under consideration the expression for the EP
can be presented as follows \cite{SVZ}:
\begin{eqnarray} \label{EPt}
V^{(1)}_{gen} &=& \frac{gH}{2\pi\beta}\sum \limits_{l= -\infty}^{+
\infty} \int\limits_{- \infty}^{+ \infty} \frac{dp_3}{2\pi}
\sum\limits_{n = 0,\sigma = \pm 1}^{\infty} \log [\beta^2(\omega^2_l +
\epsilon ^2_{n,\sigma,p_3} + \Pi(n,T,H,\sigma) )]\\ \nonumber &+&
\frac{1}{2\beta}\sum\limits_{l=-\infty}^{+\infty}\int
\limits_{-\infty}^{+\infty}\frac{d^3
p}{(2\pi)^3} log[\beta^2(\omega^2_l +
\vec{p}^2 + \Pi^0(H,T))].  \end{eqnarray} This generalized EP is
written as the sum of energies of the charged gluon field modes in the
external magnetic field ( $ \omega_l = \frac{2\pi l}{\beta}$ -
discrete imaginary energies)

\begin{equation} \label{spectr} \epsilon^2_{n,\sigma} = p^2_3 + (2n +
1 - 2\sigma) gH.  \end{equation} It includes the temperature masses
 $\Pi(H,n,\sigma, T)$ of the charged modes, which are also dependent
 on $H$, the level number $ n = 0, 1, ...$ and the spin projection
 $\sigma = \pm 1$, as well as the one of the neutral gluons,
 $\Pi^0(H,T)$.  Then, denoting the sum $\omega_l^2 + \epsilon^2$
 describing the tree level spectrum as $D^{-1}_{0}(p_3, H,T)$, let us
 rewrite the first term of Eq.(\ref{EPt}) as follows

\begin{eqnarray} \label{EPt1}
V^{(1)}_{gen} &= & \frac{gH}{2\pi\beta}
\sum\limits_{l= -\infty}^{+\infty}\int\limits_{-\infty}^{+\infty} \frac{d
p_3}{2\pi} \sum\limits_{n,\sigma} \log [\beta^2 D^{- 1}_0(p_3,H,T)]
\nonumber\\ 
&+& \frac{gH}{2\pi\beta} \sum\limits_{l = - \infty}^{+ \infty}
\int\limits_{-\infty}^{+\infty} \frac{d p_3}{2\pi} \{ \log [ 1 + (
\omega^2_l + p^2_3  - gH)^{-1} \Pi(H,T) ] + \nonumber \\
&&\sum\limits_{{n \not
= 0,\atop \sigma \not = +1} } \log [ 1 + D_0 ( \epsilon^2_n , H,T) 
\Pi(H,T,n,\sigma) ]\}.  
\end{eqnarray}

Here, the first term gives the one-loop contribution of the charged gluons,
calculated and investigated in detail in Refs.\cite{Enol}, \cite{SVZ},
\cite{Ska1} the second one is the sum of the ring diagrams of the unstable
mode $ \epsilon^2_{n=0, \sigma=+1} = p^2_3 - gH$ (as it can be easily checked
by expanding the logarithm in a series) and the last term describes the sum of
ring diagrams of the stable modes.

In general, the incorporation of the polarization operators into EP
may lead to incorrect account for the combinatoric factors of the
two-loop diagrams. In our case, to be sure in the correctness of the
obtaining results the following procedure has been applied. We
subtracted the term
\begin{equation} \label{loop}
V_s = \frac{gH}{2\pi\beta} \sum\limits_{l=-\infty}^{+\infty} \int\limits_
{-\infty}^{+\infty}\frac{dp_3}{2\pi} \sum\limits_{{n=0,\atop \sigma = \pm1}}^
{\infty} D_0(p_3, H, T) \Pi( H, n, T, \sigma)
\end{equation}
from Eq.(\ref{EPt}) that separates the contributions of the two-loop
diagrams of charged gluons from correlation corrections (the first
diagram in Fig.2).  Then, it is easy to check that the two-loop
diagrams containing the one of loops having the neutral gluon lines
(the first diagram in Fig.1) can be accounted for within the second
term of the EP (\ref{EPt}). After that the only two-loop diagram
describing the self-interaction of charged gluon fields in the vacuum
has been straightforwardly computed.

In one-loop order the neutral gluon contribution is a trivial
$H$-independent constant which can be omitted. However, these fields
are long-range states and they do give $H$-dependent EP through the
correlation corrections depending on the temperature and field. Below, only the longitudinal neutral modes are included because their Debye's masses $ \Pi^0 (H,T)$ are nonzero. The
corresponding EP is easily calculated and is given by the expression
which can be recognized from $H = 0$ case \cite{Kal}, \cite{Car}:
\begin{eqnarray} \label{Vring1} V_{ring} &=& \frac{1}{24}\Pi^{0}(H,T)
T^2 -\frac{T}{12\pi} [\Pi^{0}(H,T)]^{3/2}\nonumber \\ &+&
\frac{(\Pi^0(H,T))^2}{32\pi^2}[\log (\frac{4\pi T}{(\Pi^0)^{1/2}}) +
\frac{3}{4} - \gamma ],
\end{eqnarray}
and $\Pi^0(H,T)= \Pi^0_{00}(k=0,H,T) $ is the zero-zero component of
the neutral gluon field polarization operator calculated in the
external field at finite temperature and taken at zero momentum,
$\gamma$ is Euler's constant.
\footnote{This type of the EP was used in Ref.\cite{SVZ} to account
for the correlation corrections of charged gluons. However, the
corresponding correct expression is quite different (see
Eq.(\ref{vrch})). } The first term in Eq.(\ref{Vring1}) has the order
$\sim g^2$ in the coupling constant, the second term is of order $\sim
g^3$ and the last one - $\sim g^4$. Restricting ourselves by the order
$\sim g^3$ it will be omitted in what follows.  As usually, for $\Pi
(H,T)$ the high temperature limits of the functions have to be
substituted into Eqs. (\ref{Vring1}), (\ref{EPt1})\cite{Kal},
\cite{Car}.  As we noted above and one is able to check, this part of
the EP correctly accounts for the contributions of the two-loop
diagrams with the neutral gluon lines.

The detailed calculations of the one-loop EP $V^{(1)}(H,T)$ have been
carried out in Refs. \cite{SVZ}, \cite{Ska1}. The EP can be written in
the form:
\begin{equation} \label{V1}
V^{(1)} = \frac{1}{8\pi^2}\int \limits_{0}^{\infty}\frac{ds}{s^2}e^
{-i\mu^2 s} \sum \limits_{l=-\infty}^{+\infty} \exp (il^2\beta^2/4s)
\Big[\frac{gH \cos (2gHs)}{\sin (gHs)} - \frac{1}{s} \big],
\end{equation}
where an auxiliary mass parameter $\mu^2 -i \epsilon, \epsilon
\rightarrow 0,$ is introduced which regulates an infrared region and
plays the role of the normalization point in the field. It is useful
for analytic continuations from weak fields $gH \le \mu^2$ to the
field strengths $gH > \mu^2$ when an imaginary part of the EP is
calculated. Then it can be set to zero. The term with $l = 0$ gives
the well studied zero temperature part, while other terms describe the
statistical part. The proper time (s-representation) is used. By
integrating over $s$ we present Eq.(\ref{V1}) in the form convenient
for numeric investigations \cite{Ska1}:
\begin{eqnarray} \label{V11}
V^{(1)}(H, T) &=& V^{(1)}(H) + V^{(1)}_{\tau}(H, T),\nonumber\\
V^{(1)}(H) &=& \frac{1}{2}H^2 + \frac{(gH)^2}{4\pi^2}[\frac{11}{12} 
\log (gH/\mu^2) - i \frac{1}{8\pi} ],\nonumber\\
V^{(1)}_{\tau}(H,T)
= &-& \frac{g^2}{\pi^2}\frac{(gH)^{3/2}}{\beta}\sum\limits_{l=1}^{\infty}
[ \sum\limits_{p=0}^{\infty}
\frac{2 K_1(l\beta(gH)^{1/2}(1+2p)^{1/2})}{l} 
- \frac{\pi}{2}\frac{1}{l}Y_1(l\beta (gH)^{1/2}) \nonumber\\ 
&-& K_1(l\beta (gH)^{1/2})]
- i\frac{1}{2\pi}\frac{(gH)^{3/2}}{\beta}\sum\limits_{l=1}^{\infty} 
\frac{1}{l}J_1(l\beta (gH)^{1/2}),  
\end{eqnarray}
where $K_1(x)$, $Y_1(x)$ and $ J_1(x)$ are Bessel's functions. Remind
that $\mu$ is a subtraction point in the field $H$. The imaginary part
of the EP is signaling the vacuum instability.  To have a more
transparent expression let us calculate the high temperature limit $ T
>> (gH)^{1/2} >> \mu$ of the $V^{(1)}(H, T)$.  By means of the Mellin
transformation technique we obtain \cite{SVZ}, \cite{Ska1}:
\begin{eqnarray} \label{V1lim}
 V^{(1)}(H,T)&=& \frac{H^2}{2} + \frac{11}{48}\frac{g^2}{\pi^2}H^2 
\log \frac{T^2}{\mu^2} - \frac{1}{3}\frac{(gH)^{3/2}T}{\pi}\nonumber\\ 
&-&i\frac{(gH)^{3/2}T}{2\pi} + O(g^2H^2).
\end{eqnarray}
Note the important cancellation of the $H$-dependent logarithms
entering the vacuum and the statistical parts.

Now, let us consider the second term in Eq.(\ref{EPt1}). An elementary
integration gives
\begin{equation} \label{Vunst}
V_{unstable} = \frac{gH T}{2\pi} [\Pi(H,T,n=0,\sigma=+1) - gH ]^{1/2} +
 i\frac{(gH)^{3/2}T}{2\pi}.
\end{equation}
From Eqs.(\ref{V1lim}) and (\ref{Vunst}) it is seen that imaginary terms are
cancelled out in the total. The final EP is real if the relation
$\Pi_{unstable}(H, T) > gH $ holds. To check is it the case or not one has to
calculate both the field spontaneously generated in the vacuum and the
radiation masses of charged gluons. The former value can be immediately
calculated by differentiating Eq.(\ref{V1lim}) with respect to $ H$ and
setting the result to zero \cite{SVZ},\cite{Ska1}:
\begin{equation} \label{mag1} (gH)^{1/2}_c = \frac{g^2}{2\pi}T.
\end{equation}

To answer the question whether this vacuum is stable or not one has to
compute $\Pi(H,T,n,\sigma)$, which is the average value of the gluon
PO taken in the tree level state of the charged gluon, and consider
the effective mass squared $M^2(H_c,T)= \Pi(H_c,T,n=0,\sigma=+1) -
gH_c$. If it is positive, one has to conclude that radiation
corrections stabilize the vacuum (in one-loop order), otherwise it
remains unstable.

\section{Debye Mass of Neutral Gluons}

First, let us calculate the $H$-dependent Debye mass of the
neutral gluons. The following procedure will be applied. We calculate
the one-loop EP due to vacuum polarization of charged gluons in the
external field $H$ and some chemical potential, $\rho$, which plays
the role of an auxiliary parameter. Since the EP is the generating
functional of the one-particle irreducible Green's functions of the
field $\rho$, one has by differentiating it twice with respect to
$\rho$ and setting $\rho = 0$ to obtain the mass squared $m_{D~
neutral}^2 = \Pi_{00}(H,T, p_0 = 0)$.

The corresponding temperature dependent part of EP is
\begin{equation} \label{EPro}
 V^{(1)} = - \frac{gH}{4\pi
^2}\sum\limits_{l=1}^{
\infty}\int\limits_{0}^{\infty} \frac{ds}{s^2}\exp ( -l^2\beta^2/4s)[\frac{
1}{\sinh (gHs)}+ 2 \sinh (gHs)] \cosh (\beta l\rho).
\end{equation}
All the notations are obvious. Then, after the 
differentiations we get
\begin{equation} \label{dif}-m^2_D = \frac{\partial^2 V^{(1)}}{\partial
 \rho^2}_{\mid \rho = 0}  =
\frac{ gH}{\pi^2}\beta^2\frac{\partial}{\partial\beta^2} \sum\limits_{l=1}^
{\infty}\int\limits_{0}^{\infty}\frac{ds}{
s}\exp ( - \frac{l^2\beta^2eH}{4s})[\frac{1}{\sinh (s)} + 2
\sinh (s)].
\end{equation}
Expanding $\sinh ^{-1}s$ in series over Bernoulli's polynomials,
\begin{equation} \frac{1}{\sinh  s} = \frac{e^{-s}}{s} 
\sum\limits_{k=0}^{\infty}
\frac{B_k}{k!}(-2s)^k,
\end{equation}
and carrying out integration over $s$ in accordance with the standard
formula
\begin{equation} \label{Kn} \int\limits_0^{\infty} ds s^{n-1} 
\exp (-as -\frac{b}{s}) = 2 (\frac{b}{a})^{n/2} K_n(2\sqrt{ab}),
\end{equation}
$a,b > 0$, we obtain in the high temperature limit $(gH)^{1/2}/T <<
1$,
\begin{equation} \label{mD} - m^2_D = \frac{gH}{\pi^2}\beta^2\frac{\partial}
{\beta^2}\sum\limits_{l=1}^{\infty}[8\frac{K_1((gH)^{1/2}\beta l)}{l(gH)^
{1/2}\beta } + 4 K_0((gH)^{1/2}\beta l)+ O(\beta) ].
\end{equation}
Hence, summing up series by means of Mellin's transformation (see
 the Appendix) and differentiating with respect to $\beta^2$, one finds
 the Debye mass of neutral gluons,
\begin{equation} \label{md1} m^2_D =  \frac{2}{3}g^2 T^2 -
\frac{(g H)^{1/2}}{\pi}T - \frac{1}{4\pi^2}(g H)
+ O((gH)^2/T^2).
\end{equation}
Here, the first term is the well known temperature mass squared of gluon and
 other ones give field-dependent contributions. As it is seen, they are 
negative. This is important for what follows. It is also interesting 
that spin does not contribute to the Debye mass in the leading order.
 Substituting expression (\ref{md1}) into equation (\ref{Vring1}), we obtain 
the correlation corrections due to neutral gluons. 

\section{Ring Diagrams for Charged Gluons}

Now, we are going to calculate $\Pi_{unstable}(H, T)$ and
$\Pi(H,T,n, \sigma)$ which cannot be found from any effective
potential and require an explicit calculation of the mass operator of
the charged gluon field. This is a separate and sufficiently
complicate problem which is discussed in detail in the other
publication \cite{SS}. Here, we only adduce the necessary for our
present purpose results - the high temperature limits of
$\Pi_{unstable}(H,T):$ \begin{equation} \label{munst}
\Pi_{unstable}(H, T) = <n=0,\sigma=1\mid Re \Pi^{charged}_{\mu \nu}
\mid n=0, \sigma = 1> = 15.62 \frac{g^2}{4\pi} (gH)^{1/2} T ,
\end{equation}
 and of excited states $ \Pi(n \ne 0, \sigma \ne + 1)$,
 \begin{eqnarray}\label{mst} Re \Pi(p_4 = 0,n, p_3 = 0, H,T, \sigma =
 + 1) = \frac{g^2}{4\pi} (gH)^{1/2} T ( 15. 62 + 4 n ), \\ \nonumber
 Re\Pi(p_4 = 0,n, p_3 = 0, H,T, \sigma = - 1) = \frac{g^2}{4\pi}
 (gH)^{1/2} T ( 11.44 + 4 n ), \end{eqnarray} 
where the average values of the PO in the states of the spectrum
(\ref{spectr}) are calculated. These formulae correspond to the limit
$gH/ T^2 << 1$.  The operator contains also an imaginary part which
describes the decay of the states due to the transitions to lower
energy levels. But for the problem under consideration only the real
part is needed, since it describes the radiation mass.
 
Let us note the most important features of the expressions
(\ref{munst}),(\ref{mst}). It is seen, at $H = 0$ no screening
magnetic mass is produced in one-loop order, as it should \cite{Kal}.
Second, the masses squared of the modes are positive and act to
stabilize the spectrum of charged gluons at high temperatures. Thus,
one has to conclude that in a nonzero chromomagnetic field the charged
transversal gluons become massive at finite temperature.\footnote{This
conclusion is in obvious contradiction with the result of
Refs. \cite{Elp}, \cite{Pers} where no radiation mass for the state
with $n = 0, \sigma = + 1$ has been derived. These authors have
included the dependence of $\Pi(n=0,\sigma = +1, H,T)$ on $H$ through
the eigen states of the tree-level spectrum, only. As the operator
they used the zero field expression in the high temperature limit. But
for transversal modes at $p_3 = 0$ this is zero \cite{Kal}.}

The Debye mass of charged gluons is found to be \cite{SS}
\begin{equation} \label{dbmch}
\Pi_{00}(k_4 = 0, k_3 = 0, H,T) = \frac{2}{3} g^2 T^2 +  \frac{g^2}{4\pi}
 (gH)^{1/2} T (6 + 4 n),
\end{equation}
where again only the real part is presented. As it is seen, the
next-to-leading term is the positive growing function of $n$. It is
interesting to note that the $\Pi(p_4 = 0, p_3 = 0, n, \sigma, H, T)$
is more informative as compared the function $\Pi(p_4 = 0, \vec{p} =
0, \sigma, T)$ used at zero field. This is because it exactly accounts
for the transversal momenta described by $n$ variable: $p^2_1 + p^2_2
= (2n + 1)gH$.

Now, substituting the the expressions (\ref{munst}),(\ref{mst}),
(\ref{dbmch}) into Eq.(\ref{EPt1}) and integrating over momentum and
calculating the sums in $n$, we obtain the ring diagram contribution
of charged gluon fields. The result can be expressed in terms of the
generalized $\zeta$-function and looks as follows,
\begin{eqnarray} \label{vrch}
V_{ring}^{ch}& = &\frac{gH T}{2\pi} \big \{ \sqrt{2 gH_D}[ \zeta 
( - \frac{1}{2}, a_+ ) 
+ \zeta ( - \frac{1}{2}, a_- ) 
+ 2 \zeta 
( - \frac{1}{2}, a_D)] \nonumber \\
& -  &\sqrt{2 gH} [3 \zeta (-\frac{1}{2}, \frac{1}{2} ) + 
\zeta (- \frac{1}{2}, \frac{3}{2})] \nonumber \\
&+& (\Pi(H,T,n=0,
\sigma=+1))^{1/2}\big\},
 \end{eqnarray} 
where the first term in the first squared brackets
corresponds to the spin projection $\sigma = + 1$, the second term -
 $\sigma = - 1$ and the last one describes the part due to longitudinal
 charged gluons. The terms in the second squared brackets give the
 independent of $\Pi(H,T)$ part of eq.(\ref{EPt1}). The last term in the 
curly brackets is due to the radiation mass of the unstable mode. This 
expression is real for sufficiently high temperatures.  The notations are 
introduced: $gH_D = gH +
\frac{g^2}{2 \pi} (gH)^{1/2} T $ , $ a_- = \frac{1}{2} + \frac{g^2}{4\pi}
\frac{11,44(gH)^{1/2} T}{2 gH_D}, a_+ = \frac{1}{2} + \frac{g^2}{4
\pi}\frac{19,62 (gH)^{1/2} T}{2 gH_D}$ and $ a_D = (\frac{2}{3} g^2 T^2 + 
\frac{3g^2}{2\pi}(gH)^{1/2}T + gH )/2 gH_D $. 

Substituting the expression (\ref{md1}) into eq.(\ref{Vring1}) and gathering
all other contributions (\ref{V11}), (\ref{vrch}), we obtain the consistent
expression for the EP. To correctly account for the two-loop diagram
contributions we have to subtract from it the terms (\ref{loop}), which will
be calculated below.

\section{Two-Loop Contribution of Charged Gluons}

In this section, to complete the calculation of the EP, the
contribution of the two-loop vacuum diagram taking into account the
self-interaction of charged gluons is computed. As before, we use the
Feynmann gauge and the Furry picture for Green's functions in the
coordinate representation. If one denotes the propagator of the
charged gluons in the external field $H$ as $ iG_{\mu\nu}(x,y)$, the
contribution of the diagram to the EP can be written as
\begin{equation} \label{V2ch}
V_{ch}^{(2)} = -\frac{g^2}{2} [ G_{\mu\mu}(x,x) G_{\nu\nu}(x,x) +
G_{\mu\nu}(x,x) G_{\nu\mu}(x,x) - 2 G_{\mu\nu}(x,x) G_{\mu\nu}(x,x) ],
\end{equation}
where the Green function is \cite{Vant}
\begin{eqnarray} \label{Gref}
&&G_{\mu\nu}(x,y)= -\frac{1}{16\pi^2} \exp \{-ig\int\limits_{x}^{y} dz [ A(z)
+ \frac{1}{2} F\cdot (z-y) ]\} \\
&&\int\limits_{0}^{\infty}\frac{ds}{s^2} \exp \{-\frac{1}{2}\log 
\frac{\sinh  gFs}
{gFs} -\frac{i}{4} (x-y) \cdot gF coth gFs \cdot (x-y)\} e^{-2gFs}_{\mu\nu}.
\nonumber
\end{eqnarray}
Here, $A_{\mu}(x)$ and $F_{\mu\nu}$ are the potential and the field
strength tensor of the external field. The proper time (
s-representation) is introduced.  To incorporate temperature in
Eq. (\ref{Gref}) the method of Ref.  \cite{Cabo} will be used. After
rotation to the Euclidean space the Matsubara-Green function is
expressed in terms of the Green function at zero temperature by the
formula
\begin{equation} \label{MGref}
G(x, y; T) = \sum\limits_{n= -\infty}^{+\infty} (- 1)^{n + [x]) \lambda}
G(x - [x]\beta u, y - n\beta u),
\end{equation}
where the corresponding Green function at $T = 0$ enters the
right-hand side, $\beta = \frac{1}{T}, u$ = (0, 0, 0, 1),$ [x]$
denotes the integer part of $x_4/\beta$ and we use the parameter
$\lambda = 1$ for fermions, $\lambda = 0$ for boson and ghost fields.

Substituting Eq.(\ref{MGref}), where $G$ in the right-hand side is
given by Eq.(\ref{Gref}), into expression (\ref{V2ch}), we obtain the
contribution $V^{(2)}_{ch}(H,T)$. By calculating the trace and setting
$y = x$ in $G_{\mu\nu }(x,y,T)$ the temperature dependent part of it
can be written as follows,
\begin{equation} \label{V2t}
V^{(2)}_{ch}(H,T) = \frac{g^2}{(4\pi)^4} \sum\limits_{n=1}^{\infty} 
[ I^2_1(n,H,
\beta) + I_2(n,H,\beta) - 2 I_3(n,H,\beta) ],
\end{equation}
where the notation are introduced:
\begin{eqnarray} \label{V2tI}
I_1(n,H,\beta)&=& 2i \int\limits_{0}^{\infty} \frac{ds}{s} \frac{gH}{\sinh  
gHs}
e^{-\frac{n^2 \beta^2}{4s}} (1 + \cosh  2gHs )
\nonumber\\
I_2 - 2 I_3 &=& 2 \Bigl[ \int\limits_0^{\infty} \frac{ds}{s} 
e^{-\frac{n^2\beta^
2}{4s}} \frac{gH}{\sinh  gHs} \Bigr]^2 
\nonumber\\
&+& 2 \Bigl[ \int\limits_0^{\infty} \frac{ds}{s} e^{-\frac{n^2\beta^2}{4s}}
gH coth gHs \Bigr]^2 - 6 \Bigl[\int\limits_0^{\infty} 
\frac{ds}{s} e^{-\frac{n^2
\beta^2}{4s}} gH \Bigr]^2.
\end{eqnarray}
The zero temperature part is given by the term with $n = 0$ in the sum
over the discrete frequencies (see Ref.\cite{Cabo}). It contains
divergences and require a renormalization. The statistical part is
finite.
 
To find the high temperature limit of these expressions we first carry
out the integration over the parameter $s$, as in Sect.3. That can
easily be done by expanding $\sinh ^{-1}(gHs)$ and $coth(gHs)$ in
series over the Bernoulli polynomials and integrating over $s$ in
accordance with formula (\ref{Kn}).  The calculation of sums in $n$
can be carried out by means of the Mellin transformation technique
(see, for example, Ref.\cite{Ska1}). The necessary formulae are
adduced in the Appendix.  In this way we obtain for the leading terms
of the limit of interest:
\begin{equation} \label{V2chlim}
V_{ch}^{(2)}(H,T)_{\mid _{T \rightarrow \infty}} = - \frac{g^2}{4 \pi}
(gH)^{1/2} T^3 + O(gH T^2).
\end{equation}
The negative sign is important for what follows.

The high temperature limit of the term (\ref{loop}) is
\begin{equation} \label{loopl}
V_s(H,T)_{\mid _{T \rightarrow \infty}} = \frac{g^2}{6\sqrt{2}\pi} \zeta(
\frac{1}{2},- \frac{1}{2}) (gH)^{1/2} T^3 + O(gH T^2).
\end{equation}
It was also calculated by the Mellin transformation method.  This term
must be subtracted from the total EP and therefore gives the negative
contribution to the leading terms of the asymptotic expansion because
$\zeta(\frac{1}{2},-\frac{1} {2}) = 0.8093$. It is worth to mention
that these are the longitudinal modes that determine the high
temperature behaviour of $V^{(2)} (H,T)$. Having obtained the two-loop
corrections to the EP, one is able to investigate the spontaneous
magnetization of the vacuum at high temperature.

\section{Discussion}
The derived EP is expressed in terms of the well known
special functions.  Therefore, it can easily be investigated
numerically for any range of parameters entering. As usually, it is
convenient to introduce the dimensionless variables: the field $\phi =
(gH)^{1/2}/T$ and the EP $v(\phi,g) = V(H,T)/T^4$.  The vacuum
magnetization at high temperatures, $ T>> (gH)^{1/2},\phi \rightarrow
0$, will be investigated within the following limiting form of the EP:

 \begin{eqnarray} \label{Vlim} v^{total}(\phi,g)_{\mid \phi
\rightarrow 0} &=& \frac{\phi^4}{2 g^2} +\frac{11}{48} \frac{\phi^4}{\pi^2} 
\log (\frac{T^2}{\mu^2})  - \frac{1}{3} \frac{\phi^{3}}{\pi}
- \frac{g^2}{24\pi} \phi
\nonumber\\ 
&-& \frac{g^3}{3\sqrt{3}} \frac{1}{\phi+\frac{g^2}{2\pi}}  
-\frac{g^2}{4\pi} \phi - \frac{g^2}{6\sqrt{2}\pi} \cdot 0,8093 \phi
 + O(g^3).
\end{eqnarray}
The logarithmic term is signalling the asymptotic freedom of $g^2(T)$
at high temperatures \cite{CKS}, \cite{Ska1}. It includes explicitly
the dependence on the scale parameter $\mu$. Other terms present,
respectively, the high temperature asymptotics of the one-loop EP, the
neutral gluon and the charged gluon rings and the two-loop diagram
contributions. To obtain the term due to $V^{ch}_{ring}$ the
asymptotic expansion of the Zeta-function \cite{Eliz}
\begin{equation} \zeta(-\frac{1}{2}, a_D)_{\mid a_D \rightarrow \infty} =
 - \frac{2}{3} a_D^{3/2} + a_D^{1/2} - \frac{1}{48} a_D^{-1/2} + O(a_D^{-3/2})
\end{equation}\nonumber
has been used. Zeta-functions with $a_+, a_-$ do not contribute in leading
order.  Since we are searching for the fields $\phi$ of order greater then
$g^2$, we can omit the term $g^3$ and obtain for the condensed field
\begin{equation} \label{mf} 
(gH)^{1/2}_c = \frac{0.6}{\pi^{1/3}}\frac{}{}g^{4/3} T .
\end{equation}
Thus, we come to the conclusion that the ferromagnetic vacuum state
does exist at high temperatures. The correlation corrections increase
the field strength as compared to the value $(gH)^{1/2}_c \sim g^2 T$
generated in one-loop order.

Now, let us compare the results of investigations in
Refs. \cite{Enol}, \cite{SVZ}, \cite{Ska1} with that of the present
paper. In Ref.\cite{Ska1} the one-loop EP was calculated. In equation
(\ref{Vlim}) this corresponds to the first and third terms which
determine the vacuum magnetization of order $(gH)^{1/2}_{c} \sim
g^2T$.The authors of Ref. \cite{Enol} missed the important
cancellation of the logarithmically dependent on $H$ terms presenting
in both the vacuum and the statistical parts of the EP. Hence, their
estimate of the magnetic field strength generated at the GUT scale is
actually based on the zero temperature EP calculated by Savvidy
\cite{Sav}. As one can see, there are no dependent on $\log (gH)$
terms in Eq. (\ref{Vlim}).  Further, in Ref. \cite{Enol} the Debye
mass $ \sim g^2 T^2$ as the temperature dependent radiation one of the
transversal charged gluons has been substituted, that is an incorrect
assumption (although, in principle, it is not important from the point
of view of the result of our investigation because the radiation mass
of charged transversal modes has been found to stabilize the spectrum
at high temperature). The Debye mass for the transversal modes was
also used in Ref. \cite{SVZ}. Therein the mentioned cancellation of
the logarithms was established and in one-loop approximation the
condensed field of order $(gH)^{1/2}_c \sim g^2 T$ has been
obtained. However, the temperature mass $m^2_D$ was used twice.
First, as a heuristic factor providing the vacuum stabilization at
high temperature. Second, in the structure of the EP responsible for
the ring diagram contribution of the unstable mode. The latter is the
main point for the final result. Besides, the contribution of the
neutral gluons was missed and the dependence on the Landau level
number, $n$, has not been assumed at all. Therefore, the EP used in
Ref.\cite{SVZ} gave no possibility to calculate the correct value of
the condensed field.

The speculations on impossibility of the spontaneous vacuum
magnetization in Refs. \cite{Elp},\cite{Pers} are based on the results
of Ref.\cite{Enol} and the observations of the gluon magnetic mass at
finite temperature of order $\sim g^2 T$. Since the field at finite
temperature was identified with the zero temperature value
$(gH)^{1/2}(T=0)<< g^2 T$ whereas the spectrum of the charged
particles is formed at distances $l \sim (gH)^{-1/2}$ much longer then
the screening magnetic length the absence of the magnetization has
been claimed.  \footnote{This estimate is used in Ref.\cite{Bere}
devoted to investigation of cosmic magnetic fields in inflationary
universe. However, in connection with the results presented above some
of numeric estimations has to be corrected.}  Our EP is free from
these short comings. It includes all the relevant terms in leading
order and the temperature masses with the correct dependences on $H$
and $T$ for all modes. With these improvements made we come to the
conclusion that the ferromagnetic vacuum state is spontaneously
generated at high temperature and the field strength is found to be of
order $ (gH)_c^{1/2}\sim g^{4/3} T$.  For small $g(T)$ this is
stronger then the field determined in one-loop order.

To better understand the situation with the vacuum stability, let us
first consider the one-loop case. As the effective mass squared we
find for the condensed field $(gH)^{1/2}_c = (g^2/2\pi) T$: $ M^2(H,T)
= \Pi(H_c,T,n=0, \sigma =+1) - gH_c > 0 $. Thus, the vacuum
stabilization is observed.  But if one substitutes instead this value
the field strengths of order $(gH)^{1/2}_c \sim g^{4/3} T$ described
by equation (\ref{mf}) the effective mass squared becomes negative.
The gluon one-loop radiation mass does not stabilize the true vacuum
magnetic field.

Nevertheless, the one-loop result makes hopeful the idea to have the
stable vacuum due to radiation corrections to the charged gluon
spectrum. Naturally, to investigate this possibility the gluon
polarization operator with the correlation correction included should
be calculated. This problem requires an additional investigation.
Other interesting possibility is the formation at high temperatures of
the gluon electrostatic potential, so-called $A_0$ condensate (see
survey \cite{BBS}), which also acts as a stabilizing factor
\cite{SVZ}. To realize the latter scenario consistently the
simultaneous spontaneous generation of both the $A_0$ condensate and
the magnetic field should be investigated. If again the homogeneous
vacuum field will be found to be unstable with these improvements
made, the inhomogeneous fields of the lattice type discussed in papers
\cite{Amol}, \cite{Ska2},\cite{MDT} may be created. Since the
condensed magnetic field is strong at high temperatures, the lattice
structures having the cells of order $\sim 1/ (g^{4/3}T) << 1/(g^2 T
)$ are located inside the region where the fields are not screened by
the gluon magnetic mass.

These results may found applications in problems of cosmology, in
particular, in studying the primordial magnetic fields in the early
universe. In a few recent years the scenario of the evolution of the
universe in external hypermagnetic field \cite{Shap2}, \cite{EEK}
became popular. Our calculation of the vacuum magnetization
unambiguously determines the possibility of the presence of strong
magnetic fields in the hot universe. Other application is the high
temperature QCD. Here, one also should take into account the
generation of the chromomagnetic field in the deconfinig phase.  As a
conclusion we stress again that Savvidy's mechanism of the vacuum
magnetization does work at high temperature, although a number of
questions concerning the vacuum stability has to be investigated in
order to derive a final picture.\\

\noindent
The authors thank J. Bohacik for useful discussion. This work has been
partially supported by DFG grant No. 436 UKR 17/24/98. 

\section*{Appendix}

To calculate the the sums in $n$ in Sects. 3, 5 the
Mellin transformation was used.  It is described in detail in
Refs. \cite{Weld},\cite{Ska1}. Here, we adduce the results for
summations of series appeared in the considered cases.  In high
temperature limit $gH/T^2 << 1$ ( $\omega \rightarrow 0$ )one obtains:
\begin{equation} \label{a1}
\sum\limits_{n=1}^{\infty} \frac{1}{n} K_{1}(n \omega) = 
\frac{\zeta(2)}{\omega}
- \frac{\pi}{2} + \frac{\omega}{4} (\log \frac{4\pi}{\omega} - 
\gamma + 1/2 ) + O(\omega^2),
\end{equation}
\begin{equation} \label{a2}
\sum\limits_{n=1}^{\infty} K_0(n \omega) = \frac{1}{2} ( \gamma  - 
\log  \frac{4\pi}{\omega})
+ O(\omega),
\end{equation}
where  $\gamma$  is Euler's constant.

\end{document}